# Transverse linear and orbital angular momenta of beam waves and propagation in random media

**MIKHAIL CHARNOTSKII,\***

*Erie, CO, 80516*

*\*Corresponding author: Mikhail.Charnotskii@gmail.com*

**For paraxial propagation of scalar waves the classic electromagnetic theory definition of transverse linear (TLM) and orbital angular (OAM) momenta of the beam wave are represented in terms of the coherence function. We show in examples that neither the presence of optical vortices is necessary for the intrinsic OAM, nor does the presence of optical vortices warrant the non-zero intrinsic OAM. The OAM is analyzed for homogeneously coherent and twisted partially coherent beam waves. A twisted Gaussian beam has an intrinsic OAM with a per-unit power value that can be continuously changed by varying the twist parameters. Using the parabolic propagation equation for the coherence function, we show that both total TLM and OAM are conserved for the free-space propagation, but not for propagation in an inhomogeneous medium. In the presence of the random inhomogeneous medium, the total TLM and OAM are conserved in average, but the OAM fluctuations grow with the propagation path. This growth is slower for beams with rotation-symmetric irradiance.**

## 1. Introduction

Orbital Angular Momentum (OAM) has been one of the most popular topics of beam propagation studies in the last decade, as evidenced by several recent books [1-5] and review papers [6-12]. This interest is motivated by two major application areas: optical manipulation of small particles and free-space optical communication [13, 14]. In most cases discussed in the literature, the OAM is associated with the vortex structure of the beam wave front. These vortex structures emerge at the nulls of the irradiance and are known under the names of phase dislocations, optical singularities, optical vortices, branch points, and screw dislocations.

The common misconception that the intrinsic OAM is inevitably associated with the phase vortices is apparently based on misinterpretation of the seminal Allen et al. paper [15]. This work properly stated that vortex Laguerre-Gaussian (LG) beams, carry intrinsic OAM, but it never claimed that the presence of vortices is necessary for the OAM existence. There are many examples in the OAM literature [6, 16, 17] that show that the presence of the optical vortices is not necessary for the existence of the OAM. In [18] it was shown that coherent superposition of three phase-shifted Gaussian beams can produce non-zero intrinsic OAM. Still, in the literature, beam waves carrying the intrinsic OAM are almost inevitably associated with helically phased beams. One of the intents of this paper is to reiterate that the intrinsic OAM and the optical vortices are not tightly connected. The second is to examine evolution of OAM on propagation in the free-space and inhomogeneous medium.

In Section 2, we revisit the classic definitions of the Transverse Linear Momentum (TLM) and Orbital Angular Momentum (OAM) of a paraxial scalar wave and relate both to the wave coherence function. In Section 3, we use the coherence function representation to show several examples of coherent and partially coherent beam waves that generalize the typical helical structure of the optical vortices, and demonstrate that the intrinsic OAM and optical vortices are only loosely related. In Section 4, we discuss the TLM and OAM evolution during propagation in a free-space, inhomogeneous medium, and TLM and OAM statistics for propagation in a random inhomogeneous medium such as atmospheric turbulence.

## 2. TLM and OAM of scalar paraxial waves

Starting from the general definition of the linear and angular momentum densities of the transverse electromagnetic field [19, Chapter 6], [6, 15]

$$\boldsymbol{\mathcal{L}} = \varepsilon_0 \boldsymbol{\mathcal{E}} \times \boldsymbol{\mathcal{B}}, \boldsymbol{\mathcal{M}} = \varepsilon_0 \mathbf{R} \times (\boldsymbol{\mathcal{E}} \times \boldsymbol{\mathcal{B}}). \qquad (1)$$

For the case of a linear polarized paraxial wave propagating along $z$-direction, when the scalar complex envelope of the electric field $u(\mathbf{r}, z)$ is introduced as

$$\boldsymbol{\mathcal{E}} = \hat{x} \,\text{Re}\big[u(\mathbf{r}, z) \exp(ikz - i\omega t)\big] \qquad (2)$$

the period-averaged TLM density $\mathbf{L}(\mathbf{r}, z)$ and $z$-component of the OAM density were $M(\mathbf{r}, z)$ were calculated in [15] as

$$\begin{aligned}\mathbf{L}(\mathbf{r}, z) &= \text{Im}\big[u^*(\mathbf{r}, z) \nabla_r u(\mathbf{r}, z)\big], \\ M(\mathbf{r}, z)\hat{\mathbf{z}} &= \mathbf{r} \times \mathbf{L}(\mathbf{r}, z).\end{aligned} \qquad (3)$$

Here $\mathbf{r} = (x, y)$ is the transverse coordinate, and we dropped the inconsequential prefactor $\omega\varepsilon_0$. An equivalent, but more intuitive definition for the OAM density

$$M(r,\theta,z) = \text{Im}\left[u^*(r,\theta,z)\frac{\partial}{\partial\theta}u(r,\theta,z)\right] \quad (4)$$

was used in [18] based on the polar coordinates in the transverse plane.

Both TLM and OAM densities are related to the coherence function of the optical wave

$$\gamma(\mathbf{r},\boldsymbol{\rho},z) \equiv \overline{u\left(\mathbf{r}+\frac{\boldsymbol{\rho}}{2},z\right)u^*\left(\mathbf{r}-\frac{\boldsymbol{\rho}}{2},z\right)}. \quad (5)$$

The overbar in Eq. (5) indicates the averaging over the possible fluctuations of the radiation source. The coherence function models the "slow detector" situation when the characteristic time of the source fluctuations is much smaller than the optical detector response time, as is the case in regular daylight imaging. In the case of coherent wave or "fast detector" conditions, assumed in Eqs. (3) and (4), the coherence function is just a product of field and conjugate in two points separated by vector $\boldsymbol{\rho}$. It is straightforward to show that for the coherent case

$$\begin{aligned}\mathbf{L}(\mathbf{r},z) &= -i\nabla_{\boldsymbol{\rho}}\gamma(\mathbf{r},0,z),\\ M(\mathbf{r},z)\hat{\mathbf{z}} &= -i\mathbf{r}\times\nabla_{\boldsymbol{\rho}}\gamma(\mathbf{r},0,z).\end{aligned} \quad (6)$$

After presenting $u(\mathbf{r},z)$ in the amplitude and phase form and introducing the wave irradiance $I(\mathbf{r},z)$

$$\begin{aligned}u(\mathbf{r},z) &= A(\mathbf{r},z)\exp[i\varphi(\mathbf{r},z)], A(\mathbf{r},z) \geq 0,\\ I(\mathbf{r},z) &= A^2(\mathbf{r},z) = \gamma(\mathbf{r},0,z),\end{aligned} \quad (7)$$

the TLM and OAM densities are related to the irradiance-weighted phase gradient

$$\begin{aligned}\mathbf{L}(\mathbf{r},z) &= I(\mathbf{r},z)\nabla\varphi(\mathbf{r},z),\\ M(\mathbf{r},z)\hat{\mathbf{z}} &= I(\mathbf{r},z)\mathbf{r}\times\nabla\varphi(\mathbf{r},z).\end{aligned} \quad (8)$$

Using polar coordinates in Eq. (8), the TLM and OAM densities can be presented as

$$\begin{aligned}\mathbf{L}(r,\theta,z) &= I(r,\theta,z)\left[\frac{\partial\varphi(r,\theta,z)}{\partial r}\hat{\mathbf{r}} + \frac{1}{r}\frac{\partial\varphi(r,\theta,z)}{\partial\theta}\hat{\boldsymbol{\theta}}\right],\\ M(r,\theta,z) &= I(r,\theta,z)\frac{\partial\varphi(r,\theta,z)}{\partial\theta}.\end{aligned} \quad (9)$$

The total TLM and OAM of the beam wave are

$$\mathbf{L}(z) = \iint d^2r\mathbf{L}(\mathbf{r},z), \quad M(z) = \iint d^2rM(\mathbf{r},z). \quad (10)$$

Clearly, the total OAM of the beam depends on the location of the coordinate origin. Namely, if the OAM is measured relative to point $\mathbf{r}_C$, then

$$\begin{aligned}M_{\mathbf{r}_C}(z)\hat{\mathbf{z}} &= -i\iint d^2r(\mathbf{r}-\mathbf{r}_C)\times\nabla_{\boldsymbol{\rho}}\gamma(\mathbf{r},0,z)\\ &= M(z)\hat{\mathbf{z}} - \mathbf{r}_C\times\mathbf{L}(z).\end{aligned} \quad (11)$$

Only in the case when the total TLM is zero, the beam is said to have intrinsic OAM that is frame invariant [20]. An equivalent representation of the total TLM and OAM in terms of the first and second geometrical moments of the Wigner function was presented in [16, 17].

Equation (6) can be readily extended to the partially coherent waves' case by recognizing that it is still valid for the instantaneous random TLM, but "slow" detector measures the mean TLM over the source fluctuations. In other words, for the partially coherent waves Eq. (6), provides the mean TLM and OAM densities of a partially coherent beam wave. Note that Eq. (8) and Eq. (9) are valid for the coherent waves only.

## 3. Examples

In this Section, we present several simple examples of the TLM and OAM calculations. These examples illustrate the relation between the OAM and the optical vortices. The first three examples deal with coherent beams where the phase can be identified, and give examples of the waves with Vortex OAM (VOAM) and Asymmetry OAM (AOAM), as introduced in [17]. Partially coherent beam waves are used in the last two cases. All examples consider the field in a single plane, and we drop the z-dependence in our notations for this Section.

### A. Radial Irradiance – Angular Phase (RI-AP) beams

Consider a special coherent beam case when irradiance is rotationally symmetric, and phase depends only on the angular variable

$$u(r,\theta) = \sqrt{I(r)}e^{i\varphi(\theta)}, I(r) \geq 0. \quad (12)$$

This type of beam wave includes the Laguerre-Gaussian (LG) beams that are ubiquitous in the OAM literature. However, unlike the LG beams, the RI-AP beams do not, in general, preserve their functional form in propagation. TLM and OAM densities for this case are

$$\begin{aligned}\mathbf{L}(r,\theta) &= \frac{1}{r}I(r)\frac{d\varphi(\theta)}{d\theta}(-\hat{\mathbf{x}}\sin\theta + \hat{\mathbf{y}}\cos\theta),\\ M(r,\theta) &= I(r)\frac{d\varphi(\theta)}{d\theta}.\end{aligned} \quad (13)$$

In order to have intrinsic OAM, it is necessary that

$$\begin{aligned}\mathbf{L} &= \int_0^\infty rdr\int_{-\pi}^{\pi}d\theta\mathbf{L}(r,\theta)\\ &= \left[\int_0^\infty drI(r)\right]\int_{-\pi}^{\pi}d\theta\frac{d\varphi(\theta)}{d\theta}(-\hat{\mathbf{x}}\sin\theta + \hat{\mathbf{y}}\cos\theta) = 0.\end{aligned} \quad (14)$$

This implies that the Fourier series for $\partial\varphi(\theta)/\partial\theta$ lack the $2\pi$-periodic term. Namely,

$$\frac{\partial\varphi(\theta)}{\partial\theta} = a_0 + a_2\cos 2\theta + b_2\sin 2\theta + a_3\cos 3\theta + b_3\sin 3\theta + \ldots \quad (15)$$

In order to have a continuous field, it is necessary that

$$\exp[i\varphi(\theta+2\pi)] = \exp[i\varphi(\theta)], \quad (16)$$

and, this constraint excludes the "fractional vortex" beams [21] from consideration. Accounting for Eq. (15), the phase is presented as

$$\varphi(\theta) = l\theta + \frac{a_2}{2}\sin 2\theta - \frac{b_2}{2}\cos 2\theta + \frac{a_3}{3}\sin 3\theta - \frac{b_3}{3}\cos 3\theta + \ldots, \quad (17)$$
$$l = 0, \pm 1, \pm 2, \ldots.$$

Figure 1 shows an example of a phase front given by Eq. (17) for $l = 1$. All coefficients $a_i$ and $b_i$ are zero, besides $b_6 = 5$.

For $l \neq 0$ continuity of the complex field $u(\mathbf{r})$ at $\mathbf{r} = 0$ requires that $I(0) = 0$.

With the phase given by Eq. (17) and OAM density by Eq. (13), the total OAM of the beam is

$$M = \int_0^\infty r dr \int_0^{2\pi} d\theta I(r) \frac{d\varphi(\theta)}{d\theta} = lP, \quad (18)$$

where $P$ is the total beam power. This relationship between the OAM and total power is well known for the LG [15] Bessel-Gaussian [22, 23] and Airy beams [24], but here it is extended to the arbitrary irradiance distributions, and a phase that is not a simple helix.

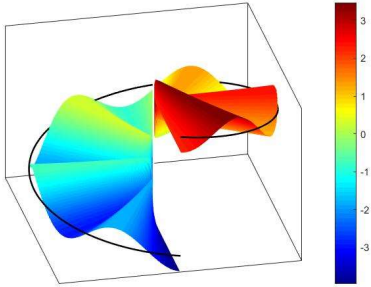

Fig. 1. Phase front described by Eq. (17) with $l = 1$ and $b_6 = 5$. Spiral curve shows trace of conventional helical phase.

### B. Vortex beam without intrinsic OAM

Consider a slightly modified version of the RI-AP beams, Eq. (12), for the field of a coherent beam where irradiance has harmonic angular dependence

$$u(r,\theta) = \sqrt{I(r)(1 + \eta \cos\theta)} e^{i\varphi(\theta)}, \; I(r) \geq 0, \; |\eta| < 1. \quad (19)$$

Using Eqs. (9) and (10), the total TLM and OAM of this beam are:

$$\mathbf{L} = \int_0^\infty I(r) dr \left[ -\hat{\mathbf{x}} \int_0^{2\pi} d\theta (1 + \eta \cos\theta) \frac{d\varphi(\theta)}{d\theta} \sin\theta \right.$$
$$\left. + \hat{\mathbf{y}} \int_0^{2\pi} d\theta (1 + \eta \cos\theta) \frac{d\varphi(\theta)}{d\theta} \cos\theta \right], \quad (20)$$
$$M = \frac{P}{2\pi} \int_0^{2\pi} d\theta (1 + \eta \cos\theta) \frac{d\varphi(\theta)}{d\theta}.$$

For this example, we keep the first three terms of the Fourier series for the phase and require that the total TLM is zero, which leads to equations

$$2b_1 + \eta b_2 = 0, \; 2\eta l + 2a_1 + \eta a_2 = 0 \quad (21)$$

The total OAM of this beam is

$$M = P\left(l + \frac{\eta}{2} a_1\right), \quad (22)$$

and the coherent beam with the field

$$u(r,\theta) = \sqrt{I(r)(1 + \eta \cos\theta)}$$
$$\exp\left[il\theta - i\frac{2l}{\eta}\sin\theta + \left(\frac{2}{\eta^2} - 1\right) l \sin 2\theta\right] \quad (23)$$

has a phase vortex but zero OAM. This is an example of VOAM and AOAM [17] canceling each other. Figure 2 shows phase and irradiance of such a beam for $l = 1$, $\eta = 0.75$ and $I(r) = r^2 \exp(-r^2)$.

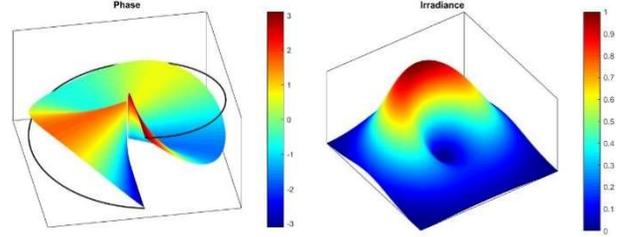

Fig. 2. Phase and irradiance of the vortex beam wave without OAM described by Eq. (23) with $l = 1$ and $\eta = 0.75$. Spiral curve shows trace of conventional spiral phase.

### C. Intrinsic OAM beams without phase vortices

Consider a coherent beam wave with irradiance and phase distributions:

$$I(r,\theta) = A^2(r)\left[1 + \eta(r^2 - 2rd\cos\theta + d^2)\right],$$
$$\text{Re}[A(r)] > 0, \; \eta > 0, \quad (24)$$
$$\varphi(r,\theta) = r(a_1 \sin\theta + b_1 \cos\theta + a_2 \sin 2\theta + b_2 \cos 2\theta)$$

This choice of the irradiance and phase distributions is motivated by the integral formula for the OAM that follows from Eq. (9) and Eq. (10)

$$M = \int_0^\infty r dr \int_0^{2\pi} d\theta I(r,\theta) \frac{\partial \varphi(r,\theta)}{\partial \theta}. \quad (25)$$

Examination of the integral over the angular variable, $\theta$, for a fixed radius $r$ suggests that the presence of the angular dependence of the irradiance distribution can possibly result in the non-zero value of this integral even for the zero-average periodic phase, as given in Eq. (24).

The irradiance in Eq. (24) is a zero-centered rotationally symmetric distribution modified by a "bump" displaced by distance $d$ along the $x$–axis. The phase is continuous across the whole plane due to the absence of the linear in $\theta$ term, compared to Eq. (17), and the presence of the $r$ factor.

Direct calculation of the integrals for TLM and OAM, Eq. (10), leads to

$$\mathbf{L} = \left[b_1(P_1 + \eta d^2 P_1 + \eta P_3) - \frac{3}{2} b_2 \eta d P_2 \right]\hat{\mathbf{x}}$$
$$+ \left[a_1(P_1 + \eta d^2 P_1 + \eta P_3) - \frac{3}{2} a_2 \eta d P_2\right]\hat{\mathbf{y}}, \quad (26)$$
$$M = -a_1 \eta d P_3.$$

Here we introduced shorthand notations for the geometrical moments of $A^2(r)$

$$P_n = 2\pi \int_0^\infty r^n dr A^2(r). \quad (27)$$

Total TLM can be set to zero by choosing

$$a_2 = \frac{2}{3\eta d P_2}\left(P_1 + \eta d^2 P_1 + \eta P_3\right)a_1,$$
$$b_2 = \frac{2}{3\eta d P_2}\left(P_1 + \eta d^2 P_1 + \eta P_3\right)b_1,$$
(28)

and intrinsic OAM does not vanish if $a_1 \neq 0$. Figures 3 and 4 show an example of such a beam wave with Gaussian $A(r) = \exp(-r^2/2w^2)$. The values of the parameters are given in the figures' captions. We emphasize that the irradiance of this beam is positive, the unwrapped phase is continuous and has no vortices or any kind of singularities, but the beam has intrinsic total OAM $M \approx -0.21P$.

Without performing the detailed calculations of the first and second moments of the Wigner function, one cannot be certain that this is an example of the pure AOAM beam. Indeed, Example E of [17] is a vortex-less beam with both VOAM and OAM components of its intrinsic OAM. An example of vortex-less beams' pure AOAM intrinsic OAM was constructed in [6] by passing an elliptical Gaussian beam through an astigmatic lens. Similar to our example, it was noted in [6] that the total OAM per unit power can be continuously varied by adjusting the beam and lens parameters, and can greatly exceed unity.

### D. Homogeneously coherent partially coherent (HCPC) beams

We now turn our attention to the more general case of partially coherent beams. Eq. (6) relates the average over the source fluctuations TLM and OAM densities to the wave coherence function. The simplest case of the partially coherent beam wave is the HCPC beam waves with coherence function

$$\gamma(\mathbf{r},\boldsymbol{\rho}) \equiv A\left(\mathbf{r}+\frac{\boldsymbol{\rho}}{2}\right)A\left(\mathbf{r}-\frac{\boldsymbol{\rho}}{2}\right)$$
$$\times \exp\left[i\varphi\left(\mathbf{r}+\frac{\boldsymbol{\rho}}{2}\right) - i\varphi\left(\mathbf{r}-\frac{\boldsymbol{\rho}}{2}\right)\right]\chi(\boldsymbol{\rho}).$$
(29)

Here

$$A(\mathbf{r}) \geq 0, \ \text{Im}[\varphi(\mathbf{r})] = 0, \ \chi(0) = 1,$$
$$\hat{\chi}(\boldsymbol{\kappa}) = \frac{1}{4\pi^2}\iint d^2\rho \chi(\boldsymbol{\rho})\exp(-i\boldsymbol{\kappa}\cdot\boldsymbol{\rho}) \geq 0,$$
(30)

and the first three constraints do not cause any loss of generality, while the last constraint on the coherence coefficient spectrum warrants the necessary Semi-Positive Definiteness (SPD) property of the coherence function. Clearly, ubiquitous Gauss-Schell Model (GSM) beams form a specific case of the HCPC class of beam waves.

As was discussed in [25], this type of wave is formed when radiation from a virtual incoherent source with spatial brightness distribution proportional to $\hat{\chi}(k\mathbf{r}/L)$ passes through an aperture with complex transmission coefficient $A(\mathbf{r})\exp[i\varphi(\mathbf{r})]$.

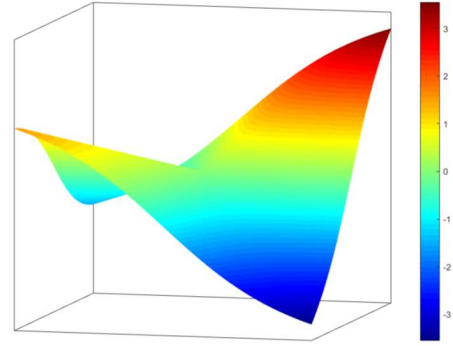

Fig. 3. Phase of the vortex-less beam wave with non-vanishing intrinsic OAM described by Eq. (24) with $a_1 = 1.0$, $b_1 = 0$, $w = 0.5$, $d = 0.5$ and $\eta = 10$.

The TLM corresponding to Eq. (29) is calculated with the help of Eq. (6) as

$$\mathbf{L}(\mathbf{r}) = A^2(\mathbf{r})\nabla\varphi(\mathbf{r}) - iI(\mathbf{r})\nabla\chi(0).$$
(31)

The first term in the right-hand part is related to the coherent effects introduced by the aperture function, and is not related to the partial coherence. It is noteworthy that the coherence coefficient enters the TLM density only as its gradient at zero separation, and the overall shape of the coherence coefficient $\chi(\mathbf{r})$ is irrelevant here. It is easy to show using the propagation model described in the next Section that $\nabla\chi(0)$ is proportional to the centroid position of the above-mentioned virtual incoherent source, and the size and brightness distribution of the source are of no importance. In fact, a coherent point source can be used with the same effect on the TLM density. The conclusion here is that the HCPC beam waves do not add much variety to the TLM and OAM in comparison to coherent waves.

### E. Twisted Gaussian beams

Simon and Mukunda [26 - 28] introduced and investigated a partially coherent twisted beam model with coherence function that can be written as

$$\gamma(\mathbf{r},\boldsymbol{\rho}) = \frac{b_0^2}{b^2}\exp\left(-\frac{R^2}{b^2} - \frac{r^2}{4\rho_c^2} + i\frac{k\sigma_\rho\sigma_\gamma t}{b^2}(\mathbf{r}\times\boldsymbol{\rho})\cdot\hat{\mathbf{z}}\right).$$
(32)

It was shown in [25] that this partially coherent beam wave can be presented as a result of averaging of the randomly displaced and tilted collimated Gaussian beam

$$u(\mathbf{r}) = \exp\left(-\frac{(\mathbf{r}-\mathbf{d})^2}{2b_0^2} + ik\boldsymbol{\gamma}\cdot(\mathbf{r}-\mathbf{d})\right).$$
(33)

Here $\mathbf{d} = (\xi,\eta)$ is the random beam displacement, and $\boldsymbol{\gamma} = (\alpha,\beta)$ is the random phase front tilt. Assuming that $\mathbf{d}$ and $\boldsymbol{\gamma}$ have zero-average normal distribution with a covariance matrix of special form

$$\left\langle \begin{pmatrix}\xi\\\eta\\\alpha\\\beta\end{pmatrix}\begin{pmatrix}\xi\\\eta\\\alpha\\\beta\end{pmatrix}^T \right\rangle = \frac{1}{2}\begin{pmatrix} \sigma_d^2 & 0 & 0 & t\sigma_d\sigma_\gamma \\ 0 & \sigma_d^2 & -t\sigma_d\sigma_\gamma & 0 \\ 0 & -t\sigma_d\sigma_\gamma & \sigma_\gamma^2 & 0 \\ t\sigma_d\sigma_\gamma & 0 & 0 & \sigma_\gamma^2 \end{pmatrix}, \quad \textbf{(34)}$$

where $|t|\leq 1$ is the correlation coefficient of displacements and tilts, it is straightforward to check that the coherence function corresponding to this wave has the form of Eq. (32) if

$$\rho_c^2 = \frac{b_0^2 b^2}{b_0^2\left(1+k^2\sigma_\gamma^2 b_0^2\right)+\sigma_\rho^2\left(1+k^2\sigma_\gamma^2 b_0^2\left(1-t^2\right)\right)}. \quad \textbf{(35)}$$

and the beam width is $b^2 = b_0^2 + \sigma_d^2$. Note that this "constructive" approach" [25] to the coherence function formation guarantees its SPD property.

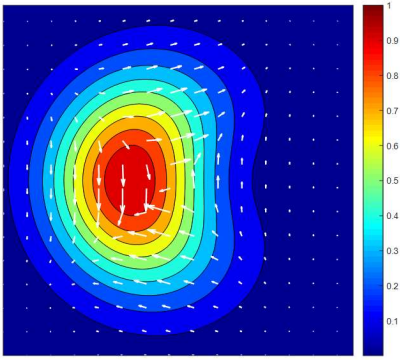

Fig. 4. Irradiance of the vortex-less beam wave with non-vanishing intrinsic OAM described by Eq. (24) with $a_1 = 1.0$, $b_1 = 0$, $w = 0.5$, $d = 0.5$ and $\eta = 10$. Arrows show the TLM vector field

Irradiance and TLM and OAM densities for coherence function, Eq. (32), can be calculated with the help of Eq. (8) as

$$I(\mathbf{r}) = \frac{b_0^2}{b^2}\exp\left(-\frac{r^2}{b^2}\right)$$
$$\mathbf{L}(\mathbf{r}) = I(\mathbf{r})\frac{k\sigma_\rho\sigma_\gamma t}{b^2}(-y\hat{\mathbf{x}} + x\hat{\mathbf{y}}), \quad \textbf{(36)}$$
$$M(\mathbf{r}) = I(\mathbf{r})\frac{k\sigma_\rho\sigma_\gamma t}{b^2}r^2.$$

Fig. 5 shows an example of the TLM vector field superimposed on the contour plot of the beam irradiance, both given by Eq. (36). The total power and intrinsic OAM of the twisted beam, Eq. (32) are

$$P = \pi b_0^2, \quad M = Pk\sigma_\rho\sigma_\gamma t. \quad \textbf{(37)}$$

It is essential to note that the individual samples of this beam, given by Eq. (33), do not carry intrinsic OAM. Namely, the total instantaneous TLM and OAM of the beam wave given by Eq. (33) are

$$\mathbf{L} = Pk\boldsymbol{\gamma}, \quad M\hat{\mathbf{z}} = Pk(\mathbf{d}\times\boldsymbol{\gamma}). \quad \textbf{(38)}$$

In the statistical averaging process, the total TLM vanishes, but the OAM is preserved to some extent by statistical correlations between the random displacement $\mathbf{d}$ and tilt $\boldsymbol{\gamma}$, which allows the intrinsic OAM to develop.

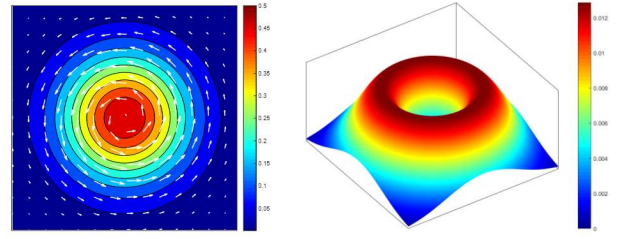

Fig. 5. Left panel – irradiance of the twisted beam described by Eq. (32) with $\sigma_d = b_0$, $k\sigma_\gamma b_0 = 0.1$, and $t = 0.7$. Arrows show the TLM vector field. Right panel – OAM density for the same beam.

### F. Intrinsic OAM of nonstationary beam wave

Discussion of the intrinsic OAM emergence for the Simon & Mukunda twisted beam suggests a simple way of generating the intrinsic OAM beams by averaging nonstationary, but not stochastic beam waves. The structure of the covariance matrix, Eq. (34), indicates that the random tilts $\boldsymbol{\gamma}$ are always orthogonal to the random displacements $\mathbf{d}$. We introduce a rotating tilted Gaussian beam:

$$u(\mathbf{r},t) = \exp\left(-\frac{(\mathbf{r}-\mathbf{r}_C(t))^2}{2b_0^2} + ik\boldsymbol{\gamma}(t)\cdot(\mathbf{r}-\mathbf{r}_C(t))\right), \quad \textbf{(39)}$$
$$\mathbf{r}_C(t) = l(\hat{\mathbf{x}}\cos t + \hat{\mathbf{y}}\sin t), \quad \boldsymbol{\gamma}(t) = \delta(-\hat{\mathbf{x}}\sin t + \hat{\mathbf{y}}\cos t).$$

The center of this beam traces a circle with radius $l$, and the phase front tilt remains orthogonal to the instantaneous radius-vector of the beam center position. For brevity we use a unit angular velocity here, but of course the actual angular rotation frequency $\omega_R$ must be much smaller than the carrier frequency. The nonstationary coherence function of this beam is

$$\gamma(\mathbf{r},\boldsymbol{\rho},t) = \exp\left(-\frac{(x-l\cos t)^2+(y-l\sin t)^2}{b_0^2}\right.$$
$$\left. -\frac{r^2}{4b_0^2} - ik\delta(\xi\sin t - \eta\cos t)\right), \quad \textbf{(40)}$$
$$\mathbf{r} = x\hat{\mathbf{x}} + y\hat{\mathbf{y}}, \quad \boldsymbol{\rho} = \xi\hat{\mathbf{x}} + \eta\hat{\mathbf{y}}.$$

The total power, TLM and OAM of this beam are readily calculated using Eq. (6) and Eq. (10) as

$$P(t) = \pi b_0^2, \quad \mathbf{L}(t) = Pk\delta(-\hat{\mathbf{x}}\sin t + \hat{\mathbf{y}}\cos t), \quad \textbf{(41)}$$
$$M(t) = Pkl\delta.$$

Period-averaged values are

$$P = \frac{1}{2\pi}\int_0^{2\pi} dt\, P(t) = \pi b_0^2, \quad \mathbf{L} = \frac{1}{2\pi}\int_0^{2\pi} dt\, \mathbf{L}(t) = 0,$$
$$M = \frac{1}{2\pi}\int_0^{2\pi} dt\, M(t) = Pkl\delta, \quad \textbf{(42)}$$

and the beam has a non-zero average intrinsic OAM. For optical manipulation applications, in a case when the particle response time is larger than the period of rotation, this nonstationary beam would exert rotational momentum just as the stationary beam with intrinsic OAM would.

## 4. Beam propagation effects on TLM and OAM

In this Section, we discuss the evolution of the TLM and OAM during paraxial beam wave propagation, first in the free space and then in the presence of an inhomogeneous medium.

### A. Free space propagation

Propagation of the paraxial beam wave field is described by the parabolic equation

$$2ik\frac{\partial u(\mathbf{r},z)}{\partial z} + \Delta_r u(\mathbf{r},z) = 0. \quad (43)$$

The corresponding equation for propagation of the coherence function is

$$ik\frac{\partial \gamma(\mathbf{r},\boldsymbol{\rho},z)}{\partial z} + \nabla_r \cdot \nabla_\rho \gamma(\mathbf{r},\boldsymbol{\rho},z) = 0, \quad (44)$$

and it extends the paraxial propagation formalism to propagation of partially coherent waves. Setting $\boldsymbol{\rho} = 0$ in Eq. (44) leads to the equation

$$\frac{\partial I(\mathbf{r},z)}{\partial z} = -\frac{1}{k}\nabla_r \cdot \mathbf{L}(\mathbf{r},z), \quad (45)$$

which, expectedly, relates the change of the power density (irradiance) to the divergence of the TLM. For a bounded beam-type wave, where the field vanishes quickly with distance from the beam axis integration of Eq. (45) over the transverse plane leads to the energy conservation principle

$$\frac{\partial}{\partial z}\iint d^2 r I(\mathbf{r},z) = \frac{\partial}{\partial z}P(z) = 0. \quad (46)$$

Total TLM can be calculated directly from the solution of Eq. (44)

$$\gamma(\mathbf{r},\boldsymbol{\rho},z) = \frac{k^2}{4\pi^2 z^2}\iint d^2 r_0 \iint d^2 \rho_0 \gamma(\mathbf{r}_0,\boldsymbol{\rho}_0,0)$$
$$\times \exp\left[\frac{ik}{z}(\mathbf{r}-\mathbf{r}_0)(\boldsymbol{\rho}-\boldsymbol{\rho}_0)\right]. \quad (47)$$

Using Eqs. (8) and (10) the total TLM after propagation from the plane $z = 0$ to plane $z$ is

$$\mathbf{L}(z) = \frac{k}{z}\frac{k^2}{4\pi^2 z^2}\iint d^2 r_0 \iint d^2 \rho_0 \gamma(\mathbf{r}_0,\boldsymbol{\rho}_0,0)$$
$$\times \iint d^2 r (\mathbf{r}-\mathbf{r}_0)\exp\left[\frac{ik}{z}(\mathbf{r}_0 - \mathbf{r})\boldsymbol{\rho}_0\right]. \quad (48)$$

The last integral in Eq. (48) is recognized as being proportional to the gradient of the Dirac delta-function $\nabla \delta(\boldsymbol{\rho}_0)$, leading to

$$\mathbf{L}(z) = -i\iint d^2 r_0 \nabla_{\rho_0}\gamma(\mathbf{r}_0,\mathbf{0},0) = \mathbf{L}(0), \quad (49)$$

and the conclusion that the total TLM of beam waves is preserved in the free-space propagation. In particular, this implies that a beam wave with intrinsic OAM in the initial plane $z = 0$ will maintain some intrinsic OAM in the propagation process.

In order to calculate the value of this OAM, following Eqs. (8) and (10), we apply $\nabla_\rho$ operator to Eq. (44), set $\boldsymbol{\rho} = 0$, cross-multiply the result by $\mathbf{r}$ and integrate over $\mathbf{r}$. The result is the equation for the total OAM change on free-space propagation

$$\frac{d}{dz}M(z) = \frac{1}{k}\iint d^2 r \hat{\mathbf{z}} \cdot \left[\mathbf{r} \times \nabla_\rho \left(\nabla_r \cdot \nabla_\rho \gamma(\mathbf{r},\boldsymbol{\rho},z)\right)\right]_{\boldsymbol{\rho}=0}. \quad (50)$$

The second integral identity, Eq. (A8), implies that the right-hand term is identically zero, and

$$M(z) = M(0). \quad (51)$$

Hence, both total TLM and OAM are conserved on the free-space propagation, and intrinsic OAM, if exists, does not change on propagation.

Conservation of OAM is widely cited in the literature, and is the foundation for application of the OAM carrying beams for free-space optical communication. However, OAM conservation is usually referred to the specific case of LG beams, or is based on the analysis of the creation and disappearance of optical vortices and associated topological charges. Here, derivation is based on the rigorous definitions simplified for the paraxial geometry, Eqs. (8) and (10), and the paraxial propagation model presented by parabolic Eq. (44). In particular, all intrinsic OAM values calculated for the coherent and partially coherent examples of vortex and vortex-less beams in the previous Section are conserved if the corresponding fields or coherence functions are used as initial conditions for wave propagation in Eqs. (43) or (44).

### B. TLM and OAM in an inhomogeneous medium

Paraxial propagation through an inhomogeneous medium with large and smooth variations of refractive index $n(\mathbf{r},z)$ is described by parabolic equation [29, 30, 31]

$$2ik\frac{\partial u(\mathbf{r},z)}{\partial z} + \Delta_r u(\mathbf{r},z) + 2k^2 n(\mathbf{r},z)u(\mathbf{r},z) = 0. \quad (52)$$

A corresponding equation for the coherence function propagation is [30]

$$ik\frac{\partial \gamma(\mathbf{r},\boldsymbol{\rho},z)}{\partial z} + \nabla_r \cdot \nabla_\rho \gamma(\mathbf{r},\boldsymbol{\rho},z)$$
$$+ k^2 \left[n\left(\mathbf{r}+\frac{\boldsymbol{\rho}}{2},z\right) - n\left(\mathbf{r}-\frac{\boldsymbol{\rho}}{2},z\right)\right]\gamma(\mathbf{r},\boldsymbol{\rho},z) = 0. \quad (53)$$

In the case of a random inhomogeneous medium, e. g. atmospheric turbulence, this equation describes averaged over the source fluctuations, but instantaneous in terms of the random medium fluctuations coherence function. Thus, it implicitly assumes that the source fluctuations are much faster than the medium fluctuations.

Setting $\boldsymbol{\rho} = 0$ in Eq. (53) leads to Eq. (45), and the integration transverse plane leads to the energy conservation principle, Eq. (46). This is a well-known result that refractive turbulence does not affect the total beam power. [30]

The equation for the TLM density can be derived by applying the $\nabla_\rho$ operator to Eq. (53) and setting $\boldsymbol{\rho} = 0$

$$\frac{\partial}{\partial z}\mathbf{L}(\mathbf{r},z) - \frac{1}{k}\nabla_\rho\left[\nabla_r \cdot \nabla_\rho \gamma(\mathbf{r},\boldsymbol{\rho},z)\right]_{\rho=0}$$
$$- k\nabla n(\mathbf{r},z)I(\mathbf{r},z) = 0. \quad (54)$$

The equation for the total TLM can be presented as

$$\frac{\partial}{\partial z}\mathbf{L}(z)-\frac{1}{k}\iint d^2 r \nabla_\rho\left[\nabla_r \cdot \nabla_\rho \gamma(\mathbf{r},\boldsymbol{\rho},z)\right]_{\rho=0}$$
$$= k\iint d^2 r \nabla n(\mathbf{r},z) I(\mathbf{r},z), \quad (55)$$

and according to the first integral identity, Eq. (A4), the second term in the left-hand part of Eq. (55) is zero. Therefore, the total TLM evolution in an inhomogeneous medium is described by the equation

$$\frac{\partial}{\partial z}\mathbf{L}(z) = k\iint d^2 r \nabla n(\mathbf{r},z) I(\mathbf{r},z). \quad (56)$$

The right-hand part of this equation is just the irradiance-weighted mean gradient of the refractive index. With reference to Eq. (8), one can speculate that Eq. (55) represents the changes of the mean phase tilt caused by the "wedge" component of inhomogeneity. Total TLM changes indicate that, in general, beam waves lose their initial intrinsic OAM immediately upon entering an inhomogeneous medium.

The equation for the OAM density can be obtained by cross-multiplying Eq. (55) by $\mathbf{r}$, and projecting the result on the $z$-axis.

$$\frac{\partial}{\partial z}M(\mathbf{r},z) - \frac{1}{k}\hat{\mathbf{z}}\left[\mathbf{r}\times\nabla_\rho\left[\nabla_r\cdot\nabla_\rho\gamma(\mathbf{r},\boldsymbol{\rho},z)\right]\right]_{\rho=0}$$
$$- k\hat{\mathbf{z}}\left[\mathbf{r}\times\nabla n(\mathbf{r},z)\right]I(\mathbf{r},z) = 0, \quad (57)$$

and the equation for the total OAM can be obtained by integrating Eq. (57) over the transverse plane:

$$\frac{\partial}{\partial z}M(z) - \frac{1}{k}\hat{\mathbf{z}}\iint d^2 r\left[\mathbf{r}\times\nabla_\rho\left[\nabla_r\cdot\nabla_\rho\gamma(\mathbf{r},\boldsymbol{\rho},z)\right]\right]_{\rho=0}$$
$$= k\hat{\mathbf{z}}\iint d^2 r\left[\mathbf{r}\times\nabla n(\mathbf{r},z)\right]I(\mathbf{r},z). \quad (58)$$

According to the second integral identity, Eq. (A8), the second term in the left-hand part of Eq. (57) is zero. Therefore, the total OAM evolution in an inhomogeneous medium is described by the equation

$$\frac{\partial}{\partial z}M(z) = k\iint d^2 r\left[x\frac{\partial n(\mathbf{r},z)}{\partial y} - y\frac{\partial n(\mathbf{r},z)}{\partial x}\right]I(\mathbf{r},z). \quad (59)$$

Just like the total TLM, the total OAM is not conserved in the inhomogeneous medium.

### C. Mean TLM and OAM in a random medium

Results of the previous Subsection are valid for instantaneous TLM and OAM values. For random inhomogeneous media such as atmospheric turbulence, it is sensible to consider statistics of the TLM and OAM densities and totals over an ensemble of possible realizations of the random field $n(\mathbf{r},z)$. The simplest statistics are the mean values $\langle \mathbf{L}(\mathbf{r},z)\rangle$, $\langle M(\mathbf{r},z)\rangle$, $\langle \mathbf{L}(z)\rangle$, and $\langle M(z)\rangle$. Here angular brackets indicate the averaging over the realizations (or time scales) of the medium fluctuations. In the case of the partially coherent sources, we assume that source fluctuations have a very short correlation time, and are not resolved by detector. The mean TLM and OAM are still related to the coherence function by Eq. (6), if the mean coherence function $\Gamma(\mathbf{r},z) = \langle \gamma(\mathbf{r},z)\rangle$ is used. The equation for the mean coherence function can be derived from Eq. (53) using the Markov approximation [32, 30]

$$\frac{\partial \Gamma(\mathbf{r},\boldsymbol{\rho},z)}{\partial z} - \frac{i}{k}\nabla_r\cdot\nabla_\rho\Gamma(\mathbf{r},\boldsymbol{\rho},z) + \frac{\pi k^2}{4}H(\boldsymbol{\rho})\Gamma(\mathbf{r},\boldsymbol{\rho},z) = 0. \quad (60)$$

Here

$$H(\boldsymbol{\rho}) = 8\iint d^2\kappa \Phi_n(\boldsymbol{\kappa},0)[1-\cos(\boldsymbol{\kappa}\boldsymbol{\rho})], \quad (61)$$

and $\Phi_n(\boldsymbol{\kappa},p)$ is a 3-D spectrum of the refractive index fluctuations. The solution of Eq. (60) is well known:

$$\Gamma(\mathbf{r},\boldsymbol{\rho},z) = \frac{k^2}{4\pi^2 z^2}\iint d^2 r_0 \iint d^2\rho_0 \gamma(\mathbf{r}_0,\boldsymbol{\rho}_0,0)$$
$$\times \exp\left(\frac{ik}{z}(\mathbf{r}-\mathbf{r}_0)(\boldsymbol{\rho}-\boldsymbol{\rho}_0) - \frac{\pi k^2}{4}\int_0^z d\zeta H\left[\boldsymbol{\rho}_0\left(1-\frac{\zeta}{z}\right) + \boldsymbol{\rho}\frac{\zeta}{z}\right]\right), \quad (62)$$

and offers a straightforward path for calculation of the mean TLM and OAM densities. We will not display these results here, but instead analyze the total mean TLM $\langle\mathbf{L}(z)\rangle$ and OAM $\langle M(z)\rangle$ based on the parabolic Eq. (60). After applying the $\nabla_\rho$ operator to Eq. (60), setting $\boldsymbol{\rho} = 0$, and integrating over $\mathbf{r}$, with the help of the first integral identity in Eq. (A2), the equation for the mean total TLM is

$$\frac{\partial}{\partial z}\langle\mathbf{L}(z)\rangle = 0. \quad (63)$$

Similarly, after additional cross-multiplication by $\mathbf{r}$, the equation for the mean total OAM is

$$\frac{\partial}{\partial z}\langle M(z)\rangle = 0, \quad (64)$$

and both total TLM and OAM are conserved in average.

### D. Fluctuations of the total TLM and OAM

As follows from Eqs. (56, 59) and Eqs. (63, 64) a random inhomogeneous medium affects the instantaneous total TLM and OAM, but not the mean TLM and OAM values. The simplest statistic that provides insight into the OAM fluctuations is the normalized variance of the total TLM and OAM

$$\sigma_{TLM}^2 = \frac{\langle L^2(z)\rangle - \langle L(z)\rangle^2}{P^2}, \quad \sigma_{OAM}^2 = \frac{\langle M^2(z)\rangle - \langle M(z)\rangle^2}{P^2}. \quad (65)$$

Using Eq. (8), the second statistical moments of TLM and OAM can be presented as

$$\langle L^2(z)\rangle = -\iint d^2 R_1 \iint d^2 R_2 \nabla_{\boldsymbol{\rho}_1}\cdot\nabla_{\boldsymbol{\rho}_2}\Gamma_4(\mathbf{R}_1,0,\mathbf{R}_2,0,z)$$
$$\langle M^2(z)\rangle = -\iint d^2 R_1 \iint d^2 R_2 (\mathbf{R}_1\times\nabla_{\boldsymbol{\rho}_1})\cdot(\mathbf{R}_2\times\nabla_{\boldsymbol{\rho}_2}) \quad (66)$$
$$\Gamma_4(\mathbf{R}_1,0,\mathbf{R}_2,0,z).$$

Here the fourth-order mean coherence function is introduced as

$$\Gamma_4(\mathbf{R}_1,\boldsymbol{\rho}_1,\mathbf{R}_2,\boldsymbol{\rho}_2,z) = \langle \gamma(\mathbf{R}_1,\boldsymbol{\rho}_1,z)\gamma(\mathbf{R}_2,\boldsymbol{\rho}_2,z)\rangle, \quad (67)$$

and propagation of the second-order instantaneous coherence function is described by Eq. (55). The propagation equation for $\Gamma_4$ is well known under the Markov approximation [32, 30]

$$\frac{\partial}{\partial z}\Gamma_4(\mathbf{R}_1,\boldsymbol{\rho}_1,\mathbf{R}_2,\boldsymbol{\rho}_2) - \frac{i}{k}\left(\nabla_{R_1}\cdot\nabla_{\rho_1} + \nabla_{R_2}\cdot\nabla_{\rho_2}\right)\Gamma_4$$

$$+ \frac{\pi k^2}{4}\Psi(\mathbf{R}_1,\boldsymbol{\rho}_1,\mathbf{R}_2,\boldsymbol{\rho}_2)\Gamma_4 = 0,$$

$$\Psi(\mathbf{R}_1,\boldsymbol{\rho}_1,\mathbf{R}_2,\boldsymbol{\rho}_2) = H(\boldsymbol{\rho}_1) + H(\boldsymbol{\rho}_2) + H\left(\mathbf{R}_1 - \mathbf{R}_2 + \frac{\boldsymbol{\rho}_1 - \boldsymbol{\rho}_2}{2}\right) \quad (68)$$

$$+ H\left(\mathbf{R}_1 - \mathbf{R}_2 - \frac{\boldsymbol{\rho}_1 - \boldsymbol{\rho}_2}{2}\right) + H\left(\mathbf{R}_1 - \mathbf{R}_2 + \frac{\boldsymbol{\rho}_1 + \boldsymbol{\rho}_2}{2}\right)$$

$$+ H\left(\mathbf{R}_1 - \mathbf{R}_2 - \frac{\boldsymbol{\rho}_1 + \boldsymbol{\rho}_2}{2}\right).$$

After applying the $\nabla_{\rho 1}$ and $\nabla_{\rho 2}$ operators to Eq. (68), setting $\boldsymbol{\rho}_{1,2} = 0$, and integrating over $\mathbf{R}_1$ and $\mathbf{R}_2$, with the help of the first integral identity in Eq. (A2), the equation for the mean square of the total TLM is

$$\frac{\partial}{\partial z}\langle L^2(z)\rangle = \frac{\pi k^2}{4}\iint d^2R_1\iint d^2R_2\langle I(\mathbf{R}_1,z)I(\mathbf{R}_2,z)\rangle \quad (69)$$
$$\times \Delta H(\mathbf{R}_1 - \mathbf{R}_2).$$

Similarly, after additional cross-multiplications by $\mathbf{R}_1$ and $\mathbf{R}_2$, the equation for the mean square of the total OAM is

$$\frac{\partial}{\partial z}\langle L^2(z)\rangle = \frac{\pi k^2}{4}\iint d^2R_1\iint d^2R_2\langle I(\mathbf{R}_1,z)I(\mathbf{R}_2,z)\rangle \quad (70)$$
$$(\mathbf{R}_1\times\nabla)(\mathbf{R}_2\times\nabla)H(\mathbf{R}_1 - \mathbf{R}_2).$$

The second moment of irradiance in Eqs. (69, 70) complicates further development, and here we consider only perturbation solutions, where the product of the free-space irradiances replaces the said second moment.

$$\langle I(\mathbf{R}_1,z)I(\mathbf{R}_2,z)\rangle \approx I_0(\mathbf{R}_1,z)I_0(\mathbf{R}_2,z). \quad (71)$$

Using the spectral representation, Eq. (61) and the normalized irradiance spectrum

$$\hat{I}_0(\boldsymbol{\kappa},z) = \frac{1}{4\pi^2 P}\iint d^2R\, I_0(\mathbf{R},z)\exp(-i\boldsymbol{\kappa}\cdot\mathbf{R}), \quad (72)$$

we present normalized (per unit power) total TLM variance as

$$\sigma_L^2(z) = 2\pi k^2\int_0^z d\zeta\iint d^2\kappa\,\kappa^2\Phi_n(\boldsymbol{\kappa},\zeta)|\hat{I}_0(\boldsymbol{\kappa},\zeta)|^2. \quad (73)$$

This equation strongly resembles the well-known equation for the variance of the beam wander, e.g. [33]

$$\sigma_C^2(z) = 2\pi\int_0^z d\zeta\left(1 - \frac{\zeta}{z}\right)^2\iint d^2\kappa\,\kappa^2\Phi_n(\boldsymbol{\kappa},\zeta)|\hat{I}_0(\boldsymbol{\kappa},\zeta)|^2, \quad (74)$$

and it can be shown that Eq. (73) is just the variance of the phase tilt or angle of arrival of the beam wave, which is not surprising, considering Eq. (8) and Eq. (9).

Similarly, normalized (per unit power) total OAM variance in the first-order of perturbation theory is

$$\sigma_M^2(z) = 2\pi k^2\int_0^z d\zeta\iint d^2\kappa\,\kappa^2\Phi_n(\boldsymbol{\kappa},\zeta)|\nabla_\kappa\hat{I}_0(\boldsymbol{\kappa},\zeta)\times\boldsymbol{\kappa}|^2. \quad (75)$$

It is easy to see that for any rotationally symmetric irradiance distribution, the cross-product in the right-hand part of Eq. (75) is zero. It can be shown that even in this case, the second-order perturbation term does not vanish, and becomes the leading term of the perturbation series for $\sigma_M^2(z)$. Note that this includes the ubiquitous case of LG beams. The details of these elaborate calculations are outside of the scope of this paper. Instead, we consider a simple example of a coherent collimated elliptical Gaussian beam with initial field distribution

$$u(\mathbf{r},0) = \exp\left(-\frac{x^2}{2a^2(0)} - \frac{y^2}{2b^2(0)}\right). \quad (76)$$

Free-space irradiance of this beam is

$$I(\mathbf{r},z) = \exp\left(-\frac{x^2}{a^2(z)} - \frac{y^2}{b^2(z)}\right),$$
$$a^2(z) = a^2(0) + \frac{z^2}{k^2 a^2(0)},\quad b^2(z) = b^2(0) + \frac{z^2}{k^2 b^2(0)}. \quad (77)$$

And the first-order $\sigma_M^2(z)$ can be readily calculated for the standard Kolmogorov turbulence spectrum

$$\Phi_n(\boldsymbol{\kappa}) = 0.033 C_n^2 \kappa^{-\frac{11}{3}} \quad (78)$$

at the plane $z = L$ from Eq. (75) as

$$\sigma_M^2(L) = \frac{0.033}{\sqrt{2}}\pi k^2\int_0^L dz\, C_n^2 |a^2(z) - b^2(z)|^{\frac{5}{6}}$$
$$\times \int_0^\infty dx\, x^{-\frac{5}{6}}\exp\left(-x\frac{a^2(z)+b^2(z)}{|a^2(z)-b^2(z)|}\right)I_1(x). \quad (79)$$

Figure 6 shows an example of the OAM variance calculated by numerical integration of Eq. (79) for turbulence conditions with a coherence radius of approximately 4 cm. The range of Fresnel numbers $N = ka^2(0)/z$ corresponding to the $a(0)$ values in Fig. (6) is $0.1 < N < 100$, and for the left side of the chart beams are divergent with $b(z) > a(z)$ for the largest part of the propagation path. For the right side of the chart, wide collimated beams propagate similarly to a plane wave, and maintain $a(z) > b(z)$. OAM variances are larger for beams with larger ellipticity in both cases. In the middle part, $N \propto 1$, and $a(z) > b(z)$ at the start, changing to $b(z) > a(z)$ closer to the end of the path and having $b(z) \approx a(z)$ in the middle. This results in the smaller average ellipticity values along the path, and smaller OAM fluctuations. Note that in our normalization, the OAM of a standard first-order LG beam is unity. Hence, for the case considered, random turbulence-induced OAM fluctuation can be comparable to the initial beam OAM, related to the original optical vortex.

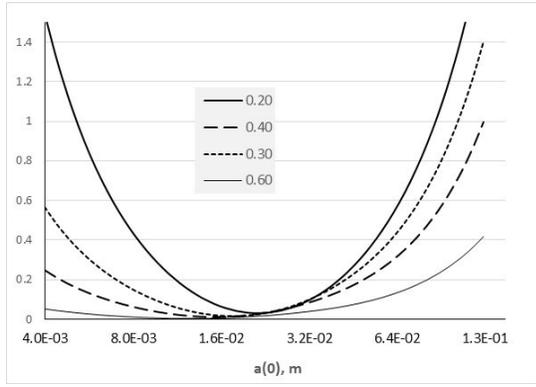

Fig. 6. Dependence of the OAM variance, $\sigma_M^2$, Eq. (79), on the initial beam size $a(0)$ for $L = 1\,km$, $C_n^2 = 10^{-14}\,m^{-2/3}$, and wavelength, $\lambda = 1.0\,\mu m$. Parameter is the ratio $b(0)/a(0)$ of the initial beam widths.

Note that both perturbation results, Eq. (73) and Eq. (75) can be extended by using the mean irradiance $\langle \hat{I}(\mathbf{\kappa}, z) \rangle$ instead of the free-space irradiance $\hat{I}_0(\mathbf{\kappa}, z)$, as was proposed for the beam wander calculations in [34].

### E. TLM and Shark-Hartman wave front sensor (WFS)

Consider a field with coherence function $\gamma(\mathbf{r}, \mathbf{\rho})$ incident on a Shack-Hartmann lenslet array. Let the $n$-th subaperture complex amplitude transmission function be

$$A(\mathbf{r} - \mathbf{R}_n)\exp\left(-i\frac{k(\mathbf{r} - \mathbf{R}_n)^2}{2F}\right),\ \text{Im}[A(\mathbf{r})] = 0. \quad (80)$$

Using Eq. (47) for the free-space propagation from the aperture plane to the focal plane, one can calculate the irradiance distribution in the focal plane $z = F$ as

$$I_n(\mathbf{r}) = \frac{k^2}{4\pi^2 F^2}\iint d^2 r_A \iint d^2 \rho_A A\left(\mathbf{r}_A + \frac{\mathbf{\rho}_A}{2}\right) A\left(\mathbf{r}_A - \frac{\mathbf{\rho}_A}{2}\right) \\ \times \gamma(\mathbf{r}_A + \mathbf{R}_n, \mathbf{\rho}_A)\exp\left[-\frac{ik}{F}\mathbf{r}\cdot\mathbf{\rho}_A\right]. \quad (81)$$

The total power in the focal plane calculated from Eq. (81) as

$$P_n = \iint d^2 r I_n(\mathbf{r}) = \iint d^2 r_A A^2(\mathbf{r}_A)\gamma(\mathbf{r}_A + \mathbf{R}_n, 0), \quad (82)$$

and is equal to the power flux through the aperture, as expected. The First Geometrical Moment (FGM) of this irradiance distribution is

$$\mathbf{G}_n = \iint d^2 r \mathbf{r} I_n(\mathbf{r}) = \frac{-iF}{k}\iint d^2 r_A A^2(\mathbf{r}_A)\nabla_{\mathbf{\rho}_A}\gamma(\mathbf{r}_A + \mathbf{R}_n, 0). \quad (83)$$

Comparing this to the first Eq. (6) reveals that the FGM is proportional to the $n$-th aperture-averaged TLM, specifically

$$\overline{\mathbf{L}}_n = \frac{1}{S_A}\iint d^2 r_A A^2(\mathbf{r}_A)\mathbf{L}(\mathbf{r}_A + R_n) = \mathbf{G}_n \frac{k}{S_A F}, \\ S_A = \iint d^2 r_A A^2(\mathbf{r}_A). \quad (84)$$

Eq. (84) shows that the FGM data obtained from the Shack-Hartmann sensor provide the information about the TLM density. Of course, as is the case with any realistic sensor, the TLM data are sub-aperture–averaged, discrete, and are limited to the wave front area intercepted by the aperture array. The OAM density and total OAM estimate can be calculated using the discrete version of the second Eq. (6) and Eq. (10):

$$M_n = \hat{\mathbf{z}}(\mathbf{R}_n \times \overline{\mathbf{L}}_n),\ M = S_A\sum_n \hat{\mathbf{z}}\cdot(\mathbf{R}_n \times \overline{\mathbf{L}}_n). \quad (85)$$

Note that the described procedure is valid for the general case of partially coherent beam waves, when the focal plane detector registers the averaged over-the-fast source fluctuations irradiance.

In comparison, the typical WFS use of the Shack-Hartmann sensor assumes a coherent incident wave, when Eq. (8) holds and the $n$-th sub-aperture FGM is

$$\mathbf{G}_n = \frac{F}{k}\iint d^2 r_A A^2(\mathbf{r}_A)I(\mathbf{r}_A + \mathbf{R}_n)\nabla\varphi(\mathbf{r}_A + \mathbf{R}_n). \quad (86)$$

The WFS strives to recover the sub-aperture–averaged phase slopes

$$\overline{\mathbf{\gamma}}_n = \frac{1}{kS_A}\iint d^2 r_A A^2(\mathbf{r}_A)\nabla\varphi(\mathbf{r}_A + \mathbf{R}_n), \quad (87)$$

and for this purpose, irradiance and power-flux fluctuations are treated as noise factors. Therefore, the mean phase-tilt estimate uses normalization by the measured power flux through the sub-aperture, $P_n$, Eq. (82):

$$\overline{\mathbf{\gamma}}_n = \frac{1}{FP_n}\mathbf{G}_n. \quad (88)$$

Phase slopes are further used to reconstruct the phase of the incident wave. In particular the "slope discrepancy" [35] can be used to detect the presence of optical vortices. However, the OAM cannot be estimated from the slopes alone.

### 5. Conclusions

For the paraxial scalar waves case, TLM and OAM densities of the electromagnetic theory are simply related to the wave coherence function in a fixed transverse plane. This allows the straightforward extension of the TLM and OAM densities concepts to the partially coherent waves case, and makes the powerful parabolic equations apparatus available for investigation of the TLM and OAM evolution on propagation.

In contrast to common beliefs, there is no definite connection between the intrinsic OAM and phase vorticity. In particular, we gave examples of single-vortex beams with zero intrinsic OAM and vortex-less beams with nonzero intrinsic OAM. In the last case, the OAM per unit power does not have discrete values. We also showed examples of the RI-AP vortex beams that have more complicated phase fronts than the simple helixes of the LG-type beams, but still have discrete values of the OAM per unit power similar to the LG beams.

We discussed the TLM and OAM of some partially coherent beam waves. For the simple case of the homogeneously coherent, or generalized Schell-type beam waves, introduction of the partial coherence does not provide any new options for creation and control of the intrinsic OAM. However, the Twisted Gaussian beam waves provide an example of a partially coherent beam with intrinsic OAM. This OAM emerges as a result of statistical averaging of instantaneous beams that do not have an intrinsic OAM.

Based on the latter example, we proposed a technique for the OAM beams generation based on the rotation of a simple coherent vortex-less tilted beam. The per-unit power OAM of these beams can be varied continuously by adjusting the rotation radius and tilt.

Based on the fundamental relation between the TLM density and coherence function, a paraxial parabolic equation can be used to study evolution of the TLM and OAM during wave propagation.

Both total TLM and OAM are conserved in the free-space propagation case. For the paraxial propagation in inhomogeneous media, in general, neither TLM nor OAM are conserved.

In the case of random inhomogeneous medium, such as atmospheric turbulence, when Markov approximation is feasible, the mean total TLM and OAM are conserved but instantaneous TLM and OAM values fluctuate.

The TLM fluctuations are closely related to the well-known effect of the wave front tilt. However, the OAM fluctuations are more delicate, and do not appear in the first-order perturbation theory for the beams with rotationally symmetric irradiance. This implies that the OAM fluctuation development will be slower for the RI-AP beams, including the LG beams, than it would be for the less symmetric beams.

We calculated the first-order perturbation term for the OAM variance of the elliptical Gaussian beams and found that for realistic atmospheric propagation conditions, the OAM fluctuations can be comparable to the intrinsic OAM of a lower-order LG beam with the same power.

Finally, we showed that the TLM density can be measured by a conventional Shack-Hartmann sensor parallel with the conventional phase slopes.

## Appendix: Two integral identities

In coordinate representation $\gamma(\mathbf{r},\boldsymbol{\rho},z) = \gamma(x,y,\xi,\eta,z)$. We introduce shorthand notations

$$f(x,y,z) = \left.\frac{\partial^2 \gamma(\mathbf{r},\boldsymbol{\rho},z)}{\partial \xi^2}\right|_{\boldsymbol{\rho}=0}, g(x,y,z) = \left.\frac{\partial^2 \gamma(\mathbf{r},\boldsymbol{\rho},z)}{\partial \xi \partial \eta}\right|_{\boldsymbol{\rho}=0},$$
$$h(x,y,z) = \left.\frac{\partial^2 \gamma(\mathbf{r},\boldsymbol{\rho},z)}{\partial \eta^2}\right|_{\boldsymbol{\rho}=0}.$$  (A1)

Consider vector

$$\mathbf{A} = \iint d^2r \nabla_\rho \left[\nabla_r \cdot \nabla_\rho \gamma(\mathbf{r},0,z)\right]_{\rho=0}$$  (A2)

having Cartesian components

$$A_x = \iint d^2r \left(\frac{\partial f}{\partial x} + \frac{\partial g}{\partial y}\right), A_y = \iint d^2r \left(\frac{\partial g}{\partial x} + \frac{\partial h}{\partial y}\right).$$  (A3)

By Green's theorem, these integral are equals to the flux of the vector fields $(f\hat{\mathbf{x}} + g\hat{\mathbf{y}})$ and $(g\hat{\mathbf{x}} + h\hat{\mathbf{y}})$ through the closed contour at infinity. For beam waves with a field quickly vanishing away from the axis, this flux is zero. This proves that

$$\mathbf{A} = \iint d^2r \nabla_\rho \left[\nabla_r \cdot \nabla_\rho \gamma(\mathbf{r},0,z)\right]_{\rho=0} = 0.$$  (A4)

Consider integral

$$B = \hat{\mathbf{z}} \iint d^2r \left[\mathbf{r} \times \nabla_\rho \left(\nabla_r \cdot \nabla_\rho \gamma(\mathbf{r},\boldsymbol{\rho},z)\right)\right]_{\boldsymbol{\rho}=0}.$$  (A5)

Using Cartesian coordinates Eq. (A5) can be written as:

$$B = \iint dxdy \left[x\left(\frac{\partial g}{\partial x} + \frac{\partial h}{\partial y}\right) - y\left(\frac{\partial f}{\partial x} + \frac{\partial g}{\partial y}\right)\right].$$  (A6)

The integrand in the right-hand part of Eq. (A1) can be presented as a divergence of auxiliary vector field $\mathbf{F}(x,y)$

$$B = \iint dxdy \nabla \cdot \mathbf{F}(x,y),$$
$$\mathbf{F}(x,y) = (-yf + xg)\hat{\mathbf{x}} + (xh - yg)\hat{\mathbf{y}}.$$  (A7)

By Green's theorem, this integral is equal to the flux of $\mathbf{F}$ through the closed contour at infinity. For beam waves with a field quickly vanishing away from the axis, this flux is zero. This proves that

$$B = \iint d^2r \left[\mathbf{r} \times \nabla_\rho \left(\nabla_r \cdot \nabla_\rho \gamma(\mathbf{r},\boldsymbol{\rho},z)\right)\right]_{\boldsymbol{\rho}=0} = 0.$$  (A8)

## References

1. M. V. Vasnetsov, K. Staliunas, eds., *Optical Vortices* (Nova Science, New York, 1999)
2. L. Allen, S. M. Barnett, and M. J. Padgett, *Optical Angular Momentum* (Optics & Optoelectronics), (CRC Press 2003)
3. A. Bekshaev, M. Soskin, and M. Vasnetsov, *Paraxial Light Beams with Angular Momentum*, (Nova Science, New York, 2008)
4. J. P. Torres and L. Torner, eds., *Twisted Photons: Applications of Light with Orbital Angular Momentum*, (Wiley, 2011)
5. D. L. Andrews, and M. Babiker, eds.,*The Angular Momentum of Light*, (Cambridge Univ. Press 2012)
6. L. Allen, M. Padgett, and M. Babiker, "The orbital angular momentum of light," Prog. Opt. **39**, 291–372 (1999).
7. M.S. Soskin and M.V. Vasnetsov, "Singular optics," Prog. Opt. **42**, 219–276 (2001).
8. A. M. Yao, and M. J. Padgett, "Orbital angular momentum: origins, behavior and applications," Adv. Opt. Phot. **3**, 161-204 (2011).
9. J. B. Gotte and S. M. Barnett, "Light beams carrying orbital angular momentum," in *The Angular Momentum of Light*, Cambridge Univ. Press (2012).
10. M. Padgett, "Light's twist," Proc. R. Soc. A **470,** 20140633 (2014)
11. A. E. Willner, H. Huang, Y. Yan, Y. Ren, N. Ahmed, G. Xie, C. Bao, L. Li, Y. Cao, Z. Zhao, J. Wang, M. P. J. Lavery, M. Tur, S. Ramachandran, A. F. Molisch, N. Ashrafi, and S. Ashrafi, "Optical communications using orbital angular momentum beams," Adv. Opt. Phot. **7**, 66–106 (2015)
12. S. M. Barnett, L. Allen, R. P. Cameron1, C. R. Gilson, M. J. Padgett, F. C. Speirits, and A. M. Yao, "On the natures of the spin and orbital parts of optical angular momentum," J. Opt. **18**, 064004 (2016)
13. G. Gibson, J. Courtial, M. J. Padgett, M. Vasnetsov, V. Pasko, S. M. Barnett, and S. Franke-Arnold, "Free-space information transfer using light beams carrying orbital angular momentum," Opt. Express **12**, 5448–5456 (2004).
14. G. Gibson, J. Courtial, M. J. Padgett, M. Vasnetsov, V. Pas'ko,S. M. Barnett, S. Franke-Arnold, "Free-space information transfer using light beams carrying orbital angular momentum," Opt. Exp. **12,** 5448-5456 (2004)
15. L. Allen, M. W. Beijersbergen, R. J. C. Spreeuw, and J. P. Woerdman," Orbital angular momentum of light and the transformation of Laguerre-Gaussian laser modes," Phys. Rev. A **45**, 8185-8189 (1992)
16. A. Ya. Bekshaev, M. S. Soskin, and M. V. Vasnetsov, ''Transformation of the orbital angular momentum of a beam with optical vortex in an astigmatic optical system,'' JETP Letters **75**, 127–130 (2002).
17. A. Ya. Bekshaev, M. V. Vasnetsov, V. G. Denisenko, and M. S. Soskin, ''Optical vortex symmetry breakdown and decomposition of the orbital angular momentum of light beams,'' J. Opt. Soc. Am. **20**, 1635-1643 (2003).
18. A. A. Kovalev, and V. V. Kotlyar, "Orbital angular momentum of superposition of identical shifted vortex beams," J. Opt. Soc. Am. A **32,** 1805-1810 (2015)
19. J. D. Jackson, *Classical Electrodynamics* 3rd ed. (Wiley, 1999).

20. A. T. O'Neil, I. MacVicar, L. Allen, and M. J. Padgett, "Intrinsic and extrinsic nature of the orbital angular momentum of a light beam," Phys. Rev. Lett. **88**, 053601(4) (2002).
21. G. Gbur, "Fractional vortex Hilbert's Hotel," Optica, 3, 222-225, )2016).
22. J. Arlt and K. Dholakia, "Generation of high-order Bessel beams by use of an axicon," Opt. Commun. **177**, 297–301 (2000).
23. I. A. Litvin, A. Dudley and A. Forbes, "Poynting vector and orbital angular momentum density of superpositions of Bessel beams," Opt. Exp. **19**, 16761 (2011)
24. H. I. Sztul and R. R. Alfano, "The Poynting vector and angular momentum of Airy beams," Opt. Exp. **16**, 9411–9416 (2008).
25. M. I. Charnotskii, "Coherence of beam arrays propagating in the turbulent atmosphere," Proc. SPIE **7685**, 76850Q-1 (2010)
26. R. Simon andN. Mukunda, "Twisted Gaussian Schell-model beams," J. Opt. Soc. Am. A **10**, 95-109 (1993).
27. R. Simon, K. Sundar, and N. Mukunda, "Twisted Gaussian Schell-model beams. I Symmetry structure and normal-mode spectrum," J. Opt. Soc. Am. A **10**, 2008-2016, (1993).
28. R. Simon, K. Sundar, and N. Mukunda, "Twisted Gaussian Schell-model beams. II Spectrum analysis and propagation characteristics," J. Opt. Soc. Am. A **10**, 2017-2023, (1993).
29. V. I. Klyatskin and V. I. Tatarskii, "The parabolic equation approximation for propagation of waves in a medium with random inhomogeneities," Sov. Phys. JETP **31**, 335–339 (1970).
30. S. M. Rytov, Yu. A. Kravtsov, and V. I. Tatarskii, *Principles of Statistical Radiophysics. 4. Wave Propagation through Random Media* (Springer, 1989).
31. M. Charnotskii, "Common omissions and and misconceptions of wave propagation in turbulence: discussion," J. Opt. Soc. Am. A **29,** 711-721 (2012).
32. V. I. Tatarskii, "Light propagation in a medium with random refractive index inhomogeneities in the Markov random process approximation," Sov. Phys. JETP **29**, 1133–1138 (1969).
33. M. Charnotskii, "Optimal beam focusing through turbulence," J. Opt. Soc. Am. A **32,** 1943-1951 (2015).
34. V. I. Klyatskin and A. I. Kon, "On the displacement of spatially bounded light beams in a turbulent medium in the Markovian-random-process approximation," Radiophys. Quantum Electron. **15**, 1056–1061 (1972).
35. T. J. Brennan and D. C. Mann, "Estimation of optical turbulence characteristics from Shack Hartmann wavefront sensor measurements," Proc. SPIE **7816** 781602 (2010)